# MVDLite: a Fast Validation Algorithm for Model View Definition Rules

Liu H.[a], Gao G.[a,b,*], Zhang H.[a,b], Liu Y.[a,b], Song Y.[c], Gu M.[a,b]
[a] Tsinghua University, China. [b] Beijing National Research Center for Information Science and Technology (BNRist), China. [c] Digital Horizon Technology Co., Ltd., China.
gaoge@tsinghua.edu.cn

**Abstract.** Model View Definition (MVD) is the standard methodology to define the data exchange requirements and rule constraints for Building Information Models (BIMs). In this paper, the MVDLite algorithm is proposed for the fast validation of MVD rules. A "rule chain" structure is introduced to combine the data templates, constraint statements, and logical interconnections in an input mvdXML ruleset, which leads to fast filtering of data nodes through the rule chain. By establishing the correspondence of each prefix of the rule chain with a string, the deep-caching strategy further improves efficiency. The outperforming experimental results show that our algorithm significantly reduces the running time of MVD validation on large real-world BIMs.

## 1. Introduction

Model View Definition (MVD) is the standard methodology to define the data exchange requirements and rule constraints for Building Information Models (BIMs). MVD defines the subsets of a specific Industry Foundation Classes (IFC) schema, with constraints on entities, attributes, geometry representations and so on.

The mvdXML is the formal representation format for MVDs recommended by buildingSMART. The mvdXML rules can be parsed by computers, which can be used for supporting software implementation in IFC-based data exchange, and for automatically validating whether IFC models conform to the MVD rules. Compared with the semantic rule-checking methods for BIMs (Pauwels, et al., 2011;2015; Beach, et al., 2015; Zhang, et al., 2019) with enriched geometry calculation and semantic inferencing, the MVD checking focuses on fast validation of data structures and values in the IFC raw data, which is recommended in the Information Delivery Manual (IDM) standard (buildingSMART, 2010).

The rules in mvdXML are represented in two separated parts: the data structures and the rule statements. The data structures are defined in the header part as nested XML tags, which represent the subgraph structures in IFC data, indicating the paths to find related data nodes starting from a root entity nodeset. The rule statements are written in "mvdXML Rule Grammar" with value constraints and logical interconnections for checking the subgraphs matched with a data template.

At present, there have been several implementations of MVD validation algorithms. In accordance with the separation of templates and rule statements in mvdXML, the current validation algorithms usually follow the two-step "matching-checking" process: first matching the template to find the subgraphs, and then checking the rule statements on each found subgraph. Different caching strategies have been applied by the algorithms, however, the efficiency of MVD validation of large rulesets on real-world size models is still a challenge.

In this paper, the MVDLite algorithm is proposed for the fast validation of MVD rules. A "rule chain" structure is introduced to combine and reorganize the data structures, value constraints and logical interconnections from an input mvdXML rule, which leads to faster searching and filtering of data nodes through the rule chain.

## 2. Related Work

### 2.1 The mvdXML Ruleset

MVD is the technical solution of the IDM in the IFC data, which intends to support meaningful IFC implementations for software developers. IDM defines data delivery process, exchange requirements and domain concepts for BIM data exchange, then MVD binds the domain concepts to IFC entities, and represents the constraints about required information for geometry, attributes and relationships in rule statements.

The mvdXML is based on several early studies about MVD rule representation. The Extended Process to Product Modeling (xPPM) (Lee, et al., 2013) is about the formal definition of IDM, and provides a tool for mapping the IDM functional parts to MVD. The Semantic Exchange Modules framework (SEM) (Venugopal, et al., 2012) is about the object-oriented definition of domain concepts in ontologies. SEM provides a mapping between the domain entity concepts and the data structure concepts in specific data forms. The Generalised Model Subset Definition (GMSD) (Weise, et al., 2003) is about the specification of rules for selecting a subset of entities from the IFC model.

Based on the previous studies about MVD, an integrated IDM-MVD process for IFC data exchange is proposed and recommended by buildingSMART (See, et al., 2012), which combines the strengths of the previous studies. The mvdXML format (Chipman, et al., 2016) is used for this integrated IDM-MVD process, which involves abundant information about exchange requirements, domain concepts, and rule constraints. An example ruleset in mvdXML is shown in Figure 1.

```
<?xml version="1.0"?>
<mvdXML xmlns:xsi="http://www.w3.org/2001/XMLSchema-instance" xmlns:xsd="http://www.w3.org/2001/XMLSchema"
uuid="ca533756-fc05-4f56-b091-04b953e362ed" status="sample" xsi:schemaLocation="http://www.buildingsmart-
tech.org/mvd/XML/1.1 http://www.buildingsmart-tech.org/mvd/XML/1.1/mvdXML_V1.1_add1.xsd"
xmlns="http://buildingsmart-tech.org/mvd/XML/1.1">
  <Templates>
    <ConceptTemplate uuid="d22c5deb-95d9-4686-800b-a23d292062c9" status="sample" applicableSchema="IFC4"
        applicableEntity="IfcWall">
      <Rules>
        <AttributeRule RuleID="IsTypedBy_0" AttributeName="IsTypedBy">
          <EntityRules>
            <EntityRule EntityName="IfcRelDefinesByType">
              <AttributeRules>
                <AttributeRule RuleID="RelatingType_1" AttributeName="RelatingType">
                  <EntityRules>
                    <EntityRule EntityName="IfcTypeObject">
                      <AttributeRules>
                        <AttributeRule RuleID="HasPropertySets_2" AttributeName="HasPropertySets">
                          <EntityRules>
                            <EntityRule EntityName="IfcPropertySet">
                              <AttributeRules>
                                <AttributeRule RuleID="Name_3" AttributeName="Name">
                                  <EntityRules>
                                    <EntityRule EntityName="IfcLabel" />
                                  </EntityRules>
                                </AttributeRule>
                                <AttributeRule RuleID="HasProperties_4" AttributeName="HasProperties">
                                  <EntityRules>
                                    <EntityRule EntityName="IfcPropertySingleValue">
                                      <AttributeRules>
                                        <AttributeRule RuleID="Name_5" AttributeName="Name">
                                          <EntityRules>
                                            <EntityRule EntityName="IfcIdentifier" />
                                          </EntityRules>
                                        </AttributeRule>
                                        <AttributeRule RuleID="NominalValue_6" AttributeName="NominalValue">
                                          <EntityRules>
                                            <EntityRule EntityName="IfcValue" />
                                          </EntityRules>
      ......

  </Templates>
  <Views>
    <ModelView uuid="d305b4eb-b668-437a-a9ef-a9ce0d55aa2b" status="sample">
      <Roots>
        <ConceptRoot uuid="9ebbc41e-b172-474a-a45e-b862ddd423b8" name="IfcWall" status="sample"
            applicableRootEntity="IfcWall">
          <Concepts>
            <Concept uuid="309fe979-2673-414a-a85b-b34b3b22fd03" name="Rule 1" status="draft" override="true">
              <Template ref="d22c5deb-95d9-4686-800b-a23d292062c9" />
              <TemplateRules operator="and">
                <TemplateRule Parameters="Name_3[Value]='Pset_WallCommon' AND Name_5[Value]='IsExternal' AND
                NominalValue_6[Type]='IfcBoolean'"/>
              </TemplateRules>
      ......

  </Views>
</mvdXML>
```

Templates about subgraph structures

Rule statements

Figure 1: An example mvdXML ruleset.

The mvdXML format involves natural language information for human-read purposes (including concept definitions and exchange requirements) as well as computer-usable rules for supporting software implementations. The computer-usable rules are in two separated parts: the data structure templates in "ConceptTemplate" tags in the header part, and the rule statements in "TemplateRules" tags in the body part. The "ConceptTemplate" part is about the subgraph structures in IFC data, indicating the paths to find related attributes and target entities starting from a certain root entity type, which is represented in nested "AttributeRule" and "EntityRule" tags. The "TemplateRules" part is about the value constraints in "mvdXML Rule Grammar" for checking the matched subgraphs, and also the applicability conditions and logical interconnections. Each attribute defined in "ConceptTemplate" is assigned with a "RuleID", so that the attribute can be referred to in the rule statements in "TemplateRules".

### 2.2 Automated MVD Validation

Automated MVD validation is a type of application to check the conformity of an IFC model with an MVD ruleset. Several studies are about modularized MVD validation (Eastman, et al., 2009; Zhang, et al., 2014; Solihin, et al., 2015; Lee, et al., 2016; 2018; Simplebim, 2022), which supports only several commonly-used MVD rule types (such as the value of attributes or the existence of relationships), and implement different program modules for different MVD validation tasks. Generalized MVD validation (Weise, et al., 2016; xBimTeam, 2016; Oraskari, et al., 2021) performs subgraph template matching and rule constraint checking directly based on the mvdXML rule, and can support arbitrary mvdXML rulesets.

Since the templates and statements are separated in mvdXML, the current MVD validation algorithms usually follow the two-step "matching-checking" process. First, for each root entity, find the subgraphs and required attributes which matches a template. Second, the rule statements (including logical combinations) are checked on each matched subgraph. The conformance result for each root entity is obtained from the existence of a subgraph that satisfies the rule statements.

For a single root entity, there are usually multiple subgraphs that can match the same template. For example, if one root entity has multiple properties and each property is an "IfcProperty" node, then for this single root entity, there are the same numbers of subgraphs that can match the template for this rule. Typically, in checking a rule on a root nodeset with $n$ entities, the total number of matched subgraphs for all the root entities is $nm$ (i.e. each root entity can match $m$ subgraphs in average), and each subgraph has $p$ attributes, then the algorithm complexity is $O(nmp)$.

Different caching strategies have been applied by the algorithms, in order that the found subgraphs can be reused in checking multiple rules. For example, one implementation in xBIM (xBimTeam, 2016) caches the found subgraphs for each root entity set in a DataTable with size $nm \times p$, and another implementation in BIMserver (Oraskari, et al., 2021) caches the edges in found subgraphs for each root entity in a HashMap with size $n \times m \times p$.

However, since the IFC model of a real-world project usually exceeds millions of nodes, with hundreds of megabytes of data, the efficiency of MVD validation of large rulesets on real-world size models is still a challenge.

## 3. The MVDLite Algorithm

The motivation of the MVDLite algorithm is based on the following two observations. First, there are usually common nodes in multiple subgraphs with different root entities. For example, if multiple instances are assigned with the same type node, then this type node (as well as properties of this type node) would be shared in the subgraphs of all instances. Second, if the constraints in rule statements are considered in subgraph matching, many false branches can be pruned at early stages in searching.

In this section, the MVDLite algorithm is proposed for the fast validation of MVD rules. First, the "rule chain" structure with several types of "rule segments" is introduced. Second, the algorithm for parsing the mvdXML rule statements and performing fast round-trip search is introduced. Then, the complexity analysis of the MVDLite algorithm is provided. Finally, the deep-caching strategy is introduced to further speed up the checking tasks.

## 3.1 The Rule Chain Structure

A rule segment is the representation of a rule as a mapping from source nodeset to target nodeset. There are three types of rule segments (attribute segment, metric segment, and compound segment). A rule chain is a sequence of rule segments starting with a root nodeset, in which the target nodeset of the former rule segment is the source nodeset of the latter rule segment.

There are three types of rule segments: attribute segment, metric segment and compound segment.

An attribute segment is a mapping defined by an attribute in the IFC data. Each attribute segment has an attribute name and a target node type, corresponding to the "AttributeRule" and "EntityRule" tags in mvdXML. Figure 2(a) shows an example attribute segment.
A metric segment is a mapping from the source nodeset to itself, which works as a filter for the source nodes. Each metric segment is with a metric (`[Type]`,`[Value]`,`[Size]`, `[Exists]`,`[Unique]`), an operator (`=, >, <, >=, <=, !=`) and a value constraint (string, boolean, or numeric value), corresponding to the components in the mvdXML Rule Grammar. Among the metric segments, the `[Type]` and `[Value]` metrics are "single metric segments", which can be evaluated by every single node in the nodeset, and act as filters for the nodeset itself. Figure 2(b) shows an example single metric segment. The `[Exists]`, `[Size]` and `[Unique]` are "collection metric segments", which can only be evaluated by a collection of nodes, and act as filters for the parent nodeset. Figure 2(c) shows an example collection metric segment.

A compound segment encapsulates one or more rule chains into brackets, and acts as a single rule segment. There are two different types of compound segments: compound attribute segment and compound metric segment. A compound attribute segment encapsulates the paths between the source nodeset and the target nodeset, and acts as a single attribute segment, as shown in Figure 3(a). A compound metric segment encapsulates the filter result of the source nodeset, and acts as a single metric segment, as shown in Figure 3(b). Each sub-chain in a compound metric segment must end with a metric segment, and each sub-chain in a compound attribute segment must end with an attribute segment. A compound segment can also be nested inside another compound segment and act as a corresponding attribute segment or metric segment.

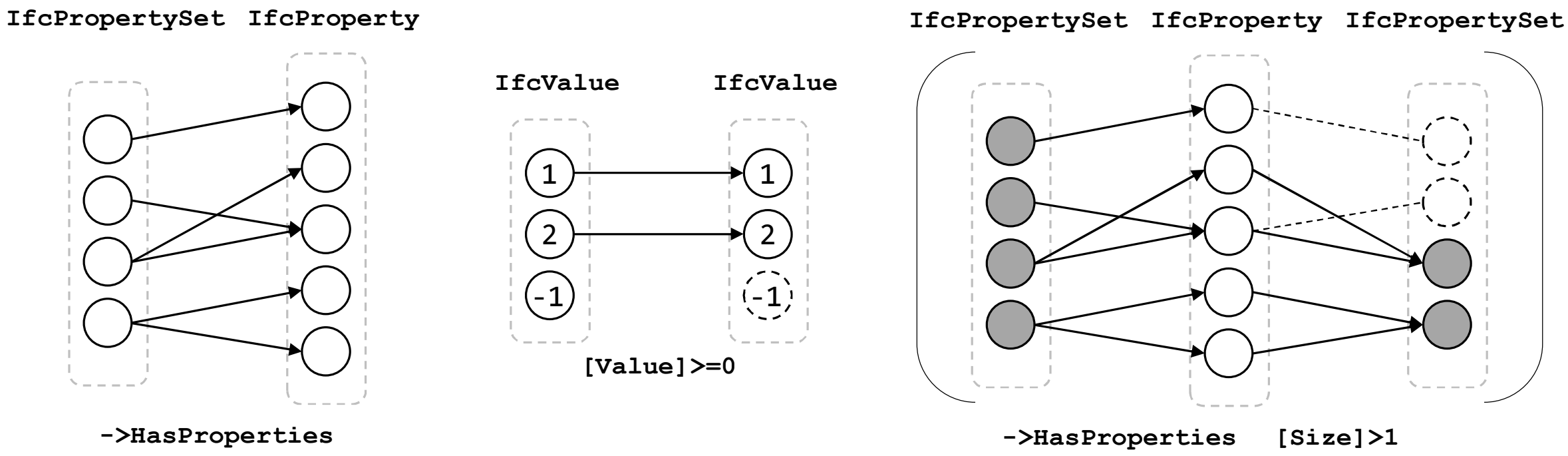

(a) Attribute segment. (b) Single metric segment. (c) Collection metric segment.

Figure 2: Examples of attribute segment and metric segments.

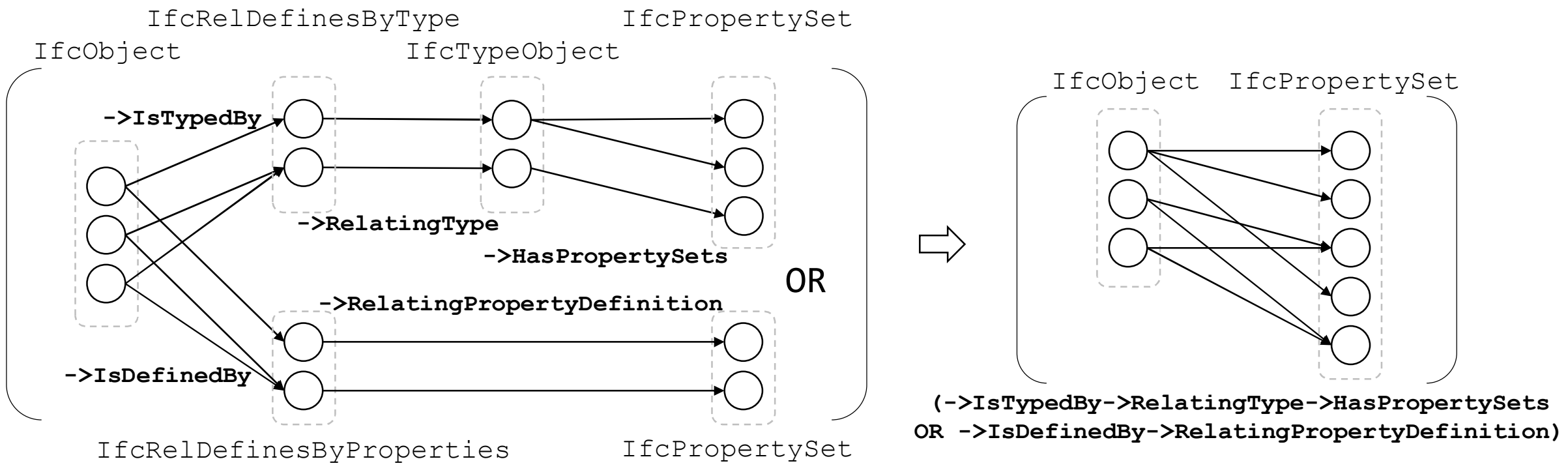

(a) Compound attribute segment.

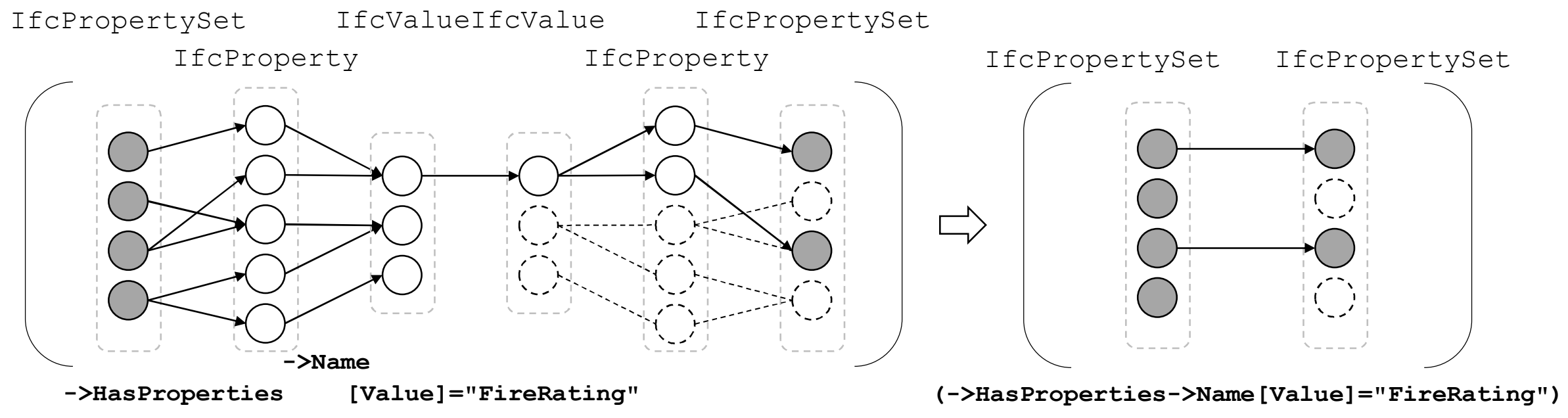

(b) Compound metric segment.

Figure 3: Examples of compound segments.

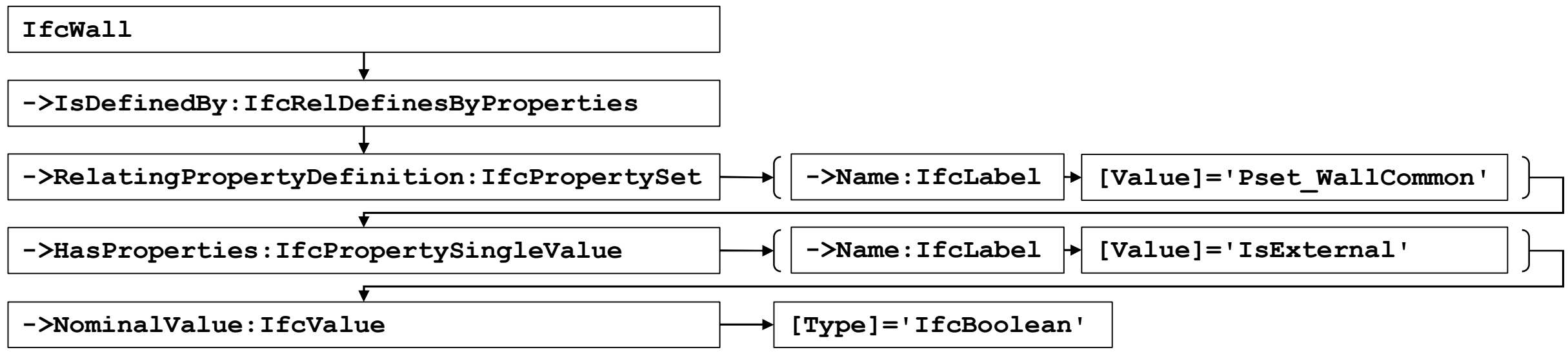

Figure 4: An example rule chain structure.

Inside a compound segment, several sub-chains can be combined with logical interconnections (`AND, OR, NOT`), which corresponds to the intersection, union and complement operations of the target nodesets. Specifically, since each metric segment can be viewed as a filter of its source nodeset, the `AND` operation of metric segments can also be represented as the series connection of metric segments.

Each rule segment is named with a string, so that each rule chain can be uniquely identified by linking the strings of all segments. An attribute segment is named with "`->`" followed by the attribute name, and the optional entity type rule can be linked afterward with a "`:`", such as "`->Name:IfcLabel`". A metric rule is named with the metric name, an operator and a value

**Algorithm 1** GeneratingRuleChain

**Input:** template **T**, rule statement **R**.
rule chain **c** = NULL
**for** (interconnection *u*, RuleID *v*, value constraint *w*) in **R** :
  **c**' = GetAttributeSegChain (*v*, **T**)
  **c**' += GetMetricSeg (*w*)
  **if c** == NULL :
    **c** = **c**'
  **else** :
    *x* = GetLowestCommonAncestor (**c**, **c**')
    **c** = MergeRuleChains(**c**, **c**', *x*, *u*)
**Output: c**

**Algorithm 2** SearchingOnRuleChain

**Input:** root nodeset $\mathbf{n}_1$, rule chain **c**, data graph **G**
**for** segment $\mathbf{s}_i$ in **c** :
  **if** $\mathbf{s}_i$ is attribute segment :
    $\mathbf{n}_{i+1}$ = FindSucceedingNodes($\mathbf{n}_i$, $\mathbf{s}_i$)
  **else if** $\mathbf{s}_i$ is metric segment :
    $\mathbf{n}_{i+1}$ = FilterNodes($\mathbf{n}_i$, $\mathbf{s}_i$)
  **else if** $\mathbf{s}_i$ is compound segment :
    $\mathbf{n}_{i+1}$ = EmptyNodeset()
    **for** (interconnection *u*, branch $\hat{\mathbf{c}}$) in $\mathbf{s}_i$ :
      $\hat{\mathbf{n}}$ = SearchingOnRuleChain($\mathbf{n}_i$, $\hat{\mathbf{c}}$, **G**)
      $\mathbf{n}_{i+1}$ = CombineNodeset($\mathbf{n}_{i+1}$, $\hat{\mathbf{n}}$, *u*)
$\mathbf{n}_{\text{out}}$ = Backtrack($\mathbf{n}_{\text{last}}$, $\mathbf{n}_1$)
**Output:** $\mathbf{n}_{\text{out}}$

constraint in order, such as "`[Value]=TRUE`". A compound segment is named with exterior brackets, inside which are the strings of the interior rule chains and logical interconnections.

Figure 4 shows a rule chain structure generated from an mvdXML rule statement. The meaning of this rule is that if a property "IsExternal" in property set "Pset_WallCommon" is assigned to an "IfcWall" instance, then the value type should be "IfcBoolean".

### 3.2 Performing Validation on the Rule Chain

The MVDLite algorithm uses mvdXML as input ruleset. The pseudo-code of the MVDLite algorithm is provided.

First, in Algorithm 1, each rule statement in an mvdXML ruleset is parsed into a rule chain structure. An mvdXML rule statement is composed of multiple sub-clauses, each with a logical interconnection, a RuleID, and a value constraint. The RuleID indicates a path from the root entity nodeset to a target nodeset, which can be parsed into a branch of the rule chain with multiple attribute segments. The value constraint can be parsed into a metric segment appended to the branch. Several branches are merged into one chain at the lowest-common-ancestor attribute. The merged chain has a shared prefix, and the suffix branches with interconnections are wrapped into compound segments.

Then, in Algorithm 2, a round-trip search is performed based on the rule chain to get the validation results. A rule chain is a series of instructions in finding and filtering the nodes. Starting from the root entity set, a round-trip search is performed: first go forward through the chain to find the existence of the paths which can pass all the rule segments, and then trace back to find the root entity nodeset where these paths started. For compound segments, each inner branch is a sub-chain starting from the current nodeset, so Algorithm 2 is performed for each branch, and the combined result acts as a single segment in the host chain.

Figure 5 compares the round-trip search algorithm on the rule chain with the "matching-checking" process in mvdXML rule validation. Rather than traversal subgraph-by-subgraph according to the template matching result, the MVDLite algorithm is able to find the results of all subgraphs through one single turn of round-trip search starting from the root nodeset.

### 3.3 The Complexity of MVDLite Algorithm

The complexity of MVD validation is mainly due to the node searching process. The complexity of the MVDLite algorithm is *O(e)*, in which *e* is the number of visited edges in the

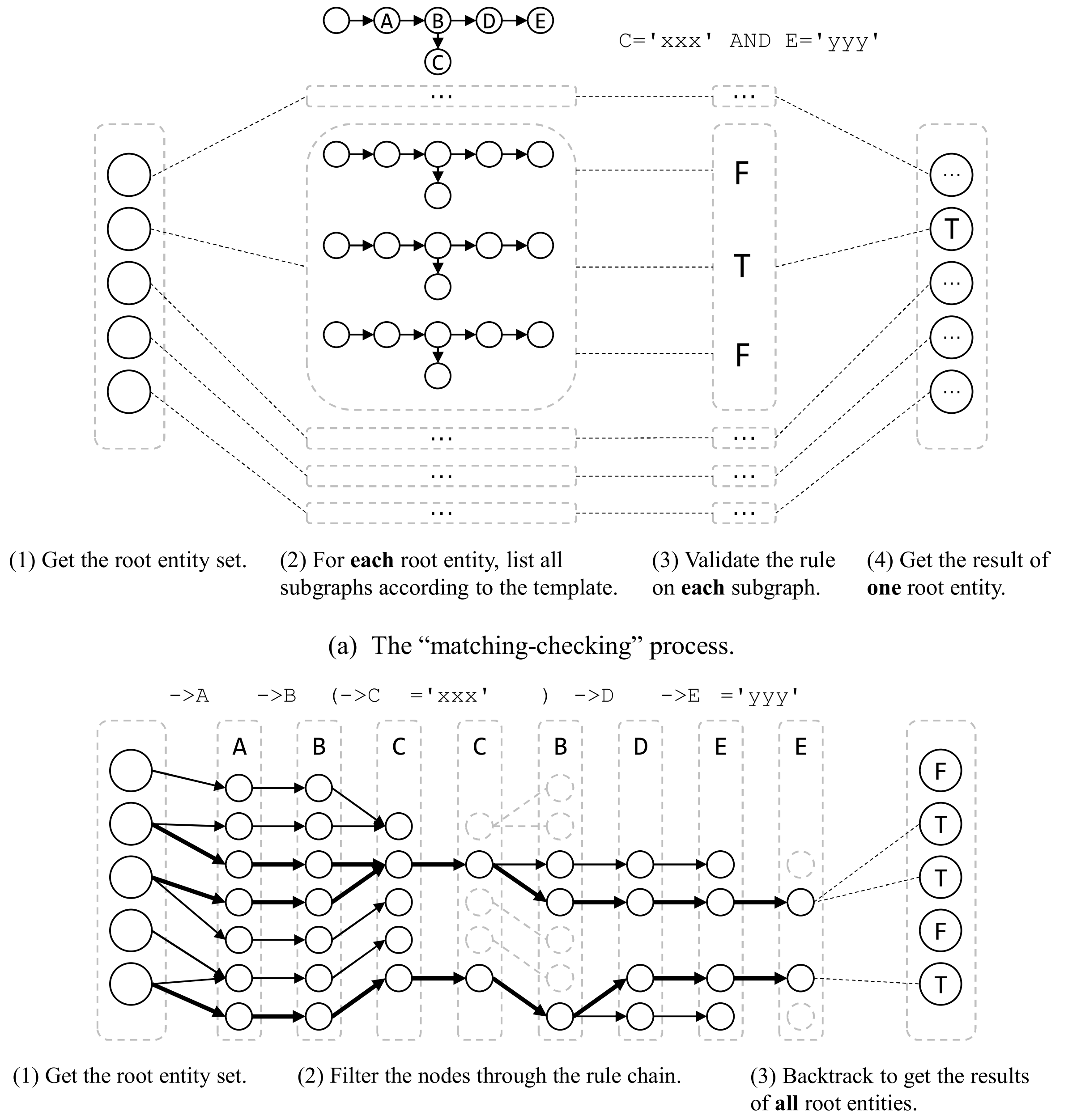

(a) The "matching-checking" process.

(b) The round-trip search on the rule chain in the MVDLite algorithm.

Figure 5: The comparison of two types of MVD rule validation algorithms.

searching, and each edge is visited at least once and at most twice (forward and backward in a round-trip). In comparison, the complexity of "matching-checking" method is *O(nmp)*.

Before comparing the complexity of the two algorithms, the formal representation of subgraph matching process in the "matching-checking" method is shown as follows.

Since the template is a directed acyclic graph, the attributes can be sorted with topological sorting, so that all prefixes of any attribute are ranked before itself. For a template with $p$ attributes (regarding the root entity set as the first attribute), let $\mathrm{f}(i)$ be the index of the direct prefix of the $i$-th attribute, $1 \leq \mathrm{f}(i) < i \leq p$. The subgraph matching process searches the nodes by each attribute, and records all matched (including partial-matched) subgraphs.

Let $\mathbf{K}^{(i)}$ be the record DataTable after scanning the $i$-th attribute, which is a matrix with the size $k_i \times i$, and $k_i$ is the number of matched subgraph prefixes. In particular, $\mathbf{K}^{(p)}$ is with size $nm \times p$, and $k_p = nm$ is the total number of subgraphs. Let $\mathbf{n}^{(i)}$ be the non-repeated nodeset of the $i$-th attribute. Specifically, the size of the root nodeset $|\mathbf{n}^{(1)}| = n$. Let $\mathbf{q}^{(i)}$ be the number of succeeding edges in scanning the $i$-th attribute for each node in $\mathbf{n}^{(\mathrm{f}(i))}$, with $|\mathbf{q}^{(i)}| = |\mathbf{n}^{(\mathrm{f}(i))}|$.

**Theorem 1.** $e \leq nmp$ .

*Proof.* In the "matching-checking" method, let $\mathbf{k}^{(i,j)}$ be the $j$-th column of $\mathbf{K}^{(i)}$, in which each item is from $\mathbf{n}^{(j)}$ and may repeat several times. Let $\mathbf{r}^{(i,j)}$ be the array of repeating times for the nodes in $\mathbf{k}^{(i,j)}$, and the size $|\mathbf{r}^{(i,j)}| = |\mathbf{n}^{(j)}|$. In scanning the $i$-th attribute, for each node in $\mathbf{k}^{(i-1,\mathrm{f}(i))}$, every succeeding edge corresponds to a new matched subgraph prefix, so

$$k_i = \mathbf{r}^{(i-1,\mathrm{f}(i))} \cdot \mathbf{q}^{(i)} . \tag{1}$$

If some subgraph prefixes do not have succeeding edges, they are kept in the DataTable as partial-matched subgraphs by appending a null node, which keeps $k_i \geq k_{i-1}$ .

In the MVDLite algorithm, since the value constraints are evaluated during the searching, some of the nodes are excluded before scanning the next attribute. Let $\mathbf{h}^{(i)}$ be a mask array for $\mathbf{n}^{(i)}$, in which kept nodes are marked with 1 and excluded nodes are marked with 0. In scanning the $i$-th attribute, the number of visited edges $e_i$ is

$$e_i = \mathbf{h}^{(\mathrm{f}(i))} \cdot \mathbf{q}^{(i)} . \tag{2}$$

Specifically, $e_1 = k_1 = n$. Since every item in $\mathbf{r}^{(i-1,\mathrm{f}(i))}$ is not less than 1, and every item in $\mathbf{h}^{(\mathrm{f}(i))}$ is not greater than 1, then $e_i \leq k_i$. As a result,

$$e = \sum_{i=1}^{p} e_i \leq \sum_{i=1}^{p} k_i \leq \sum_{i=1}^{p} k_p = nmp . \tag{3}$$

□

Theorem 1 indicates that the complexity of the MVDLite algorithm is not greater than the "matching-checking" method. The first "≤" sign gets equal only when every item in $\mathbf{r}^{(i-1,\mathrm{f}(i))}$ and $\mathbf{h}^{(\mathrm{f}(i))}$ equals 1, i.e. no repeated nodes in any subgraph prefix, and no excluded nodes by any value constraint. The second "≤" sign gets equal only when every $k_i = k_p = n$, i.e. each root entity matches only one subgraph. In real-world MVD validation tasks, the two "≤" signs usually make significant differences, which remarkably speeds up the calculation.

### 3.4 The Deep-Caching Strategy

For the checking task on a large ruleset with multiple rules, the efficiency of the MVDLite algorithm can be further improved by applying the "deep-caching" strategy. Based on the correspondence between the rule chain structure and a prefix string, the checking results can be reused across multiple rules.

By naming each rule segment with a string, the whole rule chain can be named with a string by linking the strings of all segments, and then each prefix of a rule chain has a corresponding prefix string, as shown in Figure 4. The deep-caching strategy records the correspondence so that a previously visited nodeset can be re-found according to the prefix string of a rule chain.

As a result, rather than starting from the root nodeset every time, the algorithm can start from a visited nodeset with the longest common prefix. In a large ruleset, there are usually multiple rules with a common prefix. For example, all rules for checking the type properties of an "IfcWall" instance should have the common prefix "`IfcWall->IsTypedBy:IfcRelDefinesByType->RelatingType:IfcTypeObject->HasPropertieSets:IfcPropertySet`", so the nodeset of the corresponding "IfcPropertySet" nodes can be reused across the rules. In our experiments, the deep-caching strategy is effective in speeding-up without significantly increase memory usage.

# 4. Experiments

## 4.1 Experiment Setup

The performance of the proposed MVDLite algorithm is compared with several "matching-checking" tools on models and rulesets in different sizes. The effect of the deep-caching strategy is also compared.

**Tools.** The MVDLite tool is implemented in C# based on an IFC tool named "STEParser". The compared tools include the xBIM MVD plugin (xBimTeam, 2016) with DataTable caching, and the tool based on BIMserver and IfcOpenShell (Oraskari, et al., 2021) with HashMap caching. For a fair comparison, our own "matching-checking" implementation with HashMap caching on the same STEParser tool is also tested.

**Rulesets.** Three mvdXML rulesets based on IFC4 are used, including **UnitTest.mvdxml** with 58 statements (xBimTeam, 2016), **RV.mvdxml** with 1,770 statements and **DTV.mvdxml** with 1,791 statements (buildingSMART, 2018).

**Models.** The IFC4 models used are shown in Figure 6, including two sample models from NIBS (2012) (**Duplex.ifc** with 52 MB and **Office.ifc** with 193 MB, both converted to IFC4 and merged parts in multiple disciplines), and a real-world model **B01.ifc** with 841 MB.

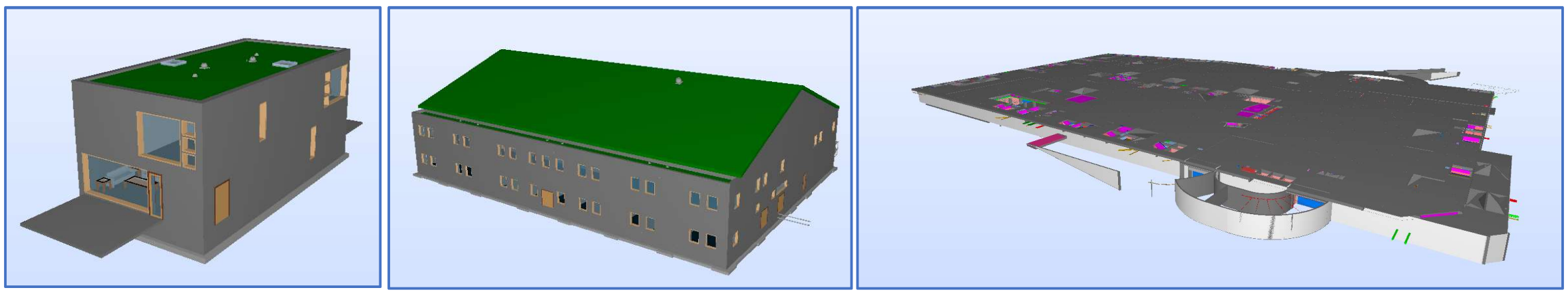

(a) Duplex.ifc (b) Office.ifc (c) B01.ifc

Figure 6: The models used in the experiments.

Table 1: Time usage in MVD validation tasks (in seconds).

| Models | Rulesets | Tools | | | | |
|---|---|---|---|---|---|---|
| | | MVDLite | MVDLite (no deep-caching) | STEParser matching-checking | (Oraskari, et al., 2021) | (xBimTeam, 2016) |
| Duplex.ifc | UnitTest.mvdxml | **0.4** | 0.5 | 1.3 | 2.5 | 2.4 |
| | RV.mvdxml | **1.3** | 2.0 | 11.2 | 21.1 | 7.1 |
| | DTV.mvdxml | **1.4** | 2.0 | 11.5 | 21.6 | 6.0 |
| Office.ifc | UnitTest.mvdxml | **0.8** | 1.5 | 5.8 | 8.6 | 177.3 |
| | RV.mvdxml | **6.0** | 9.7 | 88.5 | 97.0 | 185.7 |
| | DTV.mvdxml | **6.0** | 10.3 | 84.5 | 107.9 | 178.3 |
| B01.ifc | UnitTest.mvdxml | **5.0** | 13.1 | 49.8 | 55.8 | 10745.1 |
| | RV.mvdxml | **73.2** | 126.7 | 388.9 | 480.6 | 8722.8 |
| | DTV.mvdxml | **68.6** | 123.9 | 393.5 | 613.7 | 8707.5 |

### 4.2 Experimental Results

The time usage of the tools in MVD validation tasks are listed in Table 1 (in which the time usage for loading and parsing IFC data are excluded), with minimum values in bold. All the experiments are performed on a PC with a 3.60 GHz processor and 32GB of physical memory.

The experimental results show that the MVDLite algorithm is significantly faster than the "matching-checking" tools in all tasks. The results also show that the deep-caching strategy can improve the performance of MVD validation in checking multiple rules.

## 5. Conclusion and Future Work

The proposed MVDLite algorithm remarkably speeds up MVD validation. In a typical task, the time usage may reduce from minutes to seconds, which is beneficial in promoting IDM-MVD applications in the industry.

With the fast MVD validation algorithm, the future work will focus on the usage of MVD technology in more flexible scenarios, such as querying the IFC dataset using MVD, and extracting the partial model according to an MVD ruleset for a certain exchange requirement.

## Acknowledgements

This work was supported by the National Key Research and Development Program of China (2021YFB1600303) and the 2019 MIIT Industrial Internet Innovation and Development Project "BIM Software Industry Standardization and Public Service Platform".